\documentclass[useAMS,usenatbib]{mn2e}
\usepackage{graphicx,color,epsfig}

\newcommand{\be}{\begin{equation}}
\newcommand{\ee}{\end{equation}}

\newcommand{\apj}{ApJ}

\newcommand{\mnras}{MNRAS}
\newcommand{\aap}{A\&A}

\newcommand{\apjl}{ApJL}

\newcommand{\nat}{Nature}

\def\ltsima{$\; \buildrel < \over \sim \;$}
\def\simlt{\lower.5ex\hbox{\ltsima}}
\def\gtsima{$\; \buildrel > \over \sim \;$}
\def\simgt{\lower.5ex\hbox{\gtsima}}

\def\msun{{\,{\rm M}_\odot}}

\newcommand\mearth{{\,{\rm M}_{\oplus}}}

\def\lsun{{\,L_\odot}}
\def\del#1{{}}

\title[Tidal downsizing hypothesis]{Formation of planets by tidal downsizing
  of giant planet embryos}

\author[S. Nayakshin]{Sergei Nayakshin\\ 
Department of Physics \& Astronomy,
  University of Leicester, Leicester, LE1 7RH, UK\\
{E-mail:~} {\rm Sergei.Nayakshin@astro.le.ac.uk}}

\begin{document}

\date{Received}

\pagerange{\pageref{firstpage}--\pageref{lastpage}} \pubyear{2008}

\maketitle

\label{firstpage}

\begin{abstract}
We hypothesise that planets are made by tidal downsizing of migrating giant
planet embryos. The proposed scheme for planet formation consists of these
steps: (i) a massive young protoplanetary disc fragments at $R\sim$ several
tens to hundreds of AU on gaseous clumps with masses of a few Jupiter masses;
(ii) the clumps cool and contract, and simultaneously migrate closer in to the
parent star; (iii) as earlier suggested by \cite{Boss98}, dust sediments
inside the gas clumps to form terrestrial mass solid cores; (iv) if the solid
core becomes more massive than $\sim 10 \mearth$, a massive gas atmosphere
collapses onto the solid core; (v) when the gas clumps reach the inner few AU
from the star, tidal shear and evaporation due to stellar irradiation peel off
the outer metal-poor envelope of the clump. If tidal disruption occurs
quickly, while the system is still in stage (iii), a terrestrial planet core
is left. If it happens later, in stage (iv), a metal rich gas giant planet
with a solid core emerges from the envelope.
\end{abstract}


\section{Introduction}\label{intro}

There are currently two competing theories for planet formation in which the
planet making action starts from opposite ends.  The core accretion (CA) model
(e.g., see \cite{Safronov69,PollackEtal96,IdaLin08}, and chapters 4-6 in
\cite{Armitage10}) stipulates that planets form in a bottom-up
scenario. First, microscopic dust in the protoplanetary disc grows into
grains. There is then a poorly understood but necessary step
\citep[e.g.,][]{Wetherill90} of grains joining up somehow
\citep[see][]{YoudinGoodman05,JohansenEtal07} to make km-sized objects
(planetesimals), which then collide and stick to form ever larger solids up to
terrestrial planet masses. When the solid core reaches a critical mass of
about 10 Earth masses, a massive gas envelope builds up by accretion of gas
from the disc.

In contrast, the Gravitational Instability (GI) model suggests that giant
planets form ``big'', e.g., directly by contraction of gaseous condensations
born in a massive self-gravitating disc
\citep[e.g.,][]{Bodenheimer74,Boss97}. Giant planets in the Solar System
contain massive solid cores \citep{Fortney09}, as predicted by the core
accretion model. In the context of GI, \cite{Boss98} suggested that such cores
could result from dust sedimentation inside gaseous
proto-planets. \cite{BossEtal02} showed that removal of the gas envelope by
photoionisation due to a nearby OB star could make the ice giants of the Solar
System.

Of these two models, the GI is the one that received the lion's share of the
criticism. The most notable problems of the model are: (a) protoplanetary
discs cannot form giant planet embryos at the location of Jupiter
\citep[e.g.,][]{Rafikov05} contrary to earlier results by \cite{Boss97}; (b)
\cite{HS08,HelledEtal08} found that dust sedimentation is too slow a process
to yield observed solid cores in giant embryos of mass $\simlt$ Jupiter's
mass; (c) OB stars are too rare to explain the abundance of Jupiter-like planets observed
in extrasolar planetary systems, and (d) there seems to be no way to form
terrestrial planets like the Earth.

The goal of our Letter is to show that this criticism has long outlived
itself. All of the points raised above can be addressed if one upgrades the
1990-ies version of the GI model with more modern ideas -- formation of Giant
planet Embryos (GE hereafter) far from the parent star where the process is
allowed \citep{Rafikov05}, and then letting them migrate radially inwards, as
massive planets do \citep{GoldreichTremaine80}.

We have recently revisited (paper I, Nayakshin 2010a, to appear in MNRAS) the
suggestions of \cite{Boss98,BossEtal02} on dust sedimentation in GEs assuming
that they are formed at $R_p \sim 100$ AU. Such GEs are cooler by an order of
magnitude than those studied by Helled and co-authors. GEs with masses below
about 10-20 Jupiter masses ($M_J$) are found (paper II, Nayakshin 2010b,
submitted to MNRAS) to be excellent sites of solid core formation. Thus
addressing point (a) addresses point (b) automatically.

Below we estimate the migration time of the embryos in a realistic young
massive protoplanetary disc. We note that as embryos migrate closer to the
parent star, their Hill radius decreases faster than they can shrink due to
internal cooling. When these two are equal, we assume that the GE can be
disrupted by the tidal field of the star, which normally occurs at a few AU
distance. However, we find that grain sedimentation time is shorter than the
migration time for plausible parameter values. Thus, by the time the embryos
are due to be disrupted, they may already contain massive solid
cores. Depending on the properties of the GE at disruption, removal of the
outer metal poor gas shell reveals either a terrestrial planet or a metal-rich
giant planet. We therefore argue that such a ``tidal downsizing'' scheme may
explain planets of all types without recourse to planetesimals
 \citep[][]{Safronov69,Wetherill90}.

Our hypothesis is a hybrid scenario: giant planet formation starts with dust
sedimentation inside a self-gravitating gas clump \citep{Boss98}, as in the GI
model, but continues as in the CA model with accretion of a massive metal-rich
gas envelope from within the GE.  Additionally, solid cores left by the GE
disruption in the protoplanetary disc may grow further like in the CA model --
by accreting smaller solid debris from the disc. Giant planets may accrete
more gas and also migrate radially, as in the ``standard theory''.

\del{The only clear problem with the ``tidal
downsizing'' scheme appears that it is very rich in physics.  More
calculations detailing model outcomes need to be done for a proper comparison
with current observations of planetary systems.}

\section{Giant planet embryo model}\label{sec:estimates}
\subsection{Birth and contraction}\label{sec:embryo}


Massive young gas discs are gravitationally unstable and fragment on clumps
when the disc cooling time is no longer than a few times the local dynamical
time, $1/\Omega$ \citep[e.g.,][]{Gammie01}, where $\Omega =
(GM_*/R_p^3)^{1/2}$ is the Keplerian angular frequency at radius $R_p$ from
the star of mass $M_*$. This limits the fragmentation region to $R_p \simgt
30-100$ AU \citep[e.g.,][]{Rafikov05,SW08,Meru10}. At the point of
marginal gravitational stability the density in the disc midplane is about the
tidal density at that point, $\rho_{\rm t} = M_*/(2 \pi R_p^3)$. Thus, when
the disc just fragments, the disc cooling time is about the free-fall
timescale for the material in the disc, $t_{\rm ff} \sim 1/\sqrt{G\rho}$. As
the clumps contract further, their cooling time becomes longer than their
free-fall times. The initial state of a GE is thus that of a polytropic clump
\citep{CameronEtal82}, similar to the ``first cores'' studied in the context
of star formation \citep{Larson69,Masunaga00}. For simplicity we assume that GEs evolve
at constant mass after their formation.

Introducing dust opacity as a power-law in temperature and assuming that the
embryo is optically thick, we obtained an approximate analytic solution for
the radiative cooling of the giant embryo in paper I1. For example, for
$\kappa(T) = \kappa_0 (T/10K)$, we have
\begin{equation}
T\left(t\right) = T_0 \left[1 + 2 \frac{t}{t_0}\right]^{1/2} = 146\; m_1^{4/3}
\kappa_*^{-4/9} \left[1 + 2 \frac{t}{t_0}\right]^{1/2}\;,
\label{T_of_t}
\end{equation}
where $T_0$ is the temperature of the embryo at formation, with $m_1 = M_{\rm
  emb}/10M_J$ being the mass of the gas clump in units of $10 M_J$, $k_* =
\kappa_0/0.01$, and $t_0$ is the initial cooling time of the clump:
\begin{equation}
t_0 = 380 \; \hbox{years}\; m_1^{2/3} \kappa_{*}^{1/9}\;.
\label{tcool1}
\end{equation}
Note that the cooling time of the embryo increases with time, so that $t_{\rm
  cool}(t) \propto t$ when $t\gg t_0$. At a constant mass the embryo radius
decreases with time as
\begin{equation}
R_{\rm emb} = \frac{R_0}{ \left[1 + 2 t/t_0\right]^{1/2}} = 6.0 \;\hbox{AU}\;
\frac{\kappa_{*}^{4/9}}{m_1^{1/3}}  \left[1 + 2 \frac{t}{t_0}\right]^{-1/2}\;.
\label{r_emb}
\end{equation}
At large $t$, independently of $M_{\rm emb}$, all GEs contract as
\begin{equation}
R_{\rm emb}\left(t\right) = 0.8 \;\hbox{AU}\; \kappa_{*}^{1/2} \left[
  \frac{10^4\;\hbox{yrs}}{t}\right]^{1/2}\;.
\label{r_ass}
\end{equation}

\subsection{Dust growth and sedimentation}\label{sec:dust}

We echo the approach of \cite{Boss98} here; the resulting grain growth time scale
in a {\em constant density} embryo, for a grain with initial size $a_0$ to
reach the final size $a$ is
\begin{equation}
t_{\rm gr,0} = \frac{3 c_s}{\pi f_g \rho_{0} G R_0}\; \ln\frac{a}{a_0}\;,
\label{tgr0}
\end{equation}
where $\rho_0 = 3M_{\rm emb}/4\pi R_{\rm emb}^3$, $c_s$ and $f_g\approx 0.01$
are the initial embryo mean density, the sound speed, and the grain fraction
by mass, respectively. In paper I we improved on this model by allowing the
embryo to contract, in which case the growth time is a geometric combination
of the cooling and the initial grain growth time, $t_{\rm gr} = t_0^{3/7}
t_{\rm gr,0}^{4/7}$:
\begin{equation}
t_{\rm gr} = 3 \times 10^3 \;\hbox{yr}\;m_1^{-2/7}f_{-2}^{-4/7} k_{*}^{9/21} \left(\frac{\ln(a/a_0)}{20}\right)^{4/7}\;,
\label{tgr}
\end{equation}
where $f_{-2} = f_g/0.01$. The vaporisation time, $t_{\rm vap}$, is the time
it takes for the embryo to heat up to the vaporisation temperature of $T_{\rm
  vap}\approx 1400$ K. This is trivially obtained by solving the equation
$T(t_{\rm vap}) = T_{\rm vap}$ (using eq. \ref{T_of_t}), and a good
approximation is
\begin{equation}
t_{\rm vap} = 1.5 \times 10^4 \; \hbox{years}\; \left(\frac{T_{\rm vap}}{1400\,\hbox{K}}\right)^2 m_1^{-2} \kappa_{*}\;,
\label{tvap}
\end{equation}
Grains sediment to the centre of the core when $t_{\rm gr} < t_{\rm vap}$, and
vaporise before they could sediment if $t_{\rm gr} > t_{\rm vap}$.

The mass of refractory material (e.g., silicates) that can be used to build a
massive core is expected to be about $\sim 1/3$ of the total heavy elements
mass, or about $20$ Earth masses for $M_{\rm emb} = 10 M_J$
\citep{BossEtal02}. However, due to energy release associated with the solid
core accretion, strong convective motions near the core may develop. The
grains also melt at too high core accretion rates. Thus the final mass of the
core is usually much smaller than the maximum condensible mass (paper II).


\subsection{Radial migration of embryos}\label{sec:migration}

Gravitational torques between the disc and massive planets are significant
\citep[e.g.,][]{GoldreichTremaine80,Tanaka02}. While GEs are smaller than
their Hill's radii (see below), we assume that they migrate at similar rates.
Under this assumption, embryos migrate via type I regime if $M_{\rm emb} \le
M_t = 2 M_* (H/R)^3$ and via type II regime if $M_{\rm emb} > M_t$, where
$M_t$ is the transition mass \citep{BateEtal03}. For a self-gravitating, $Q=1$
disc, the type II migration time is
\begin{equation}
t_{\rm II}\left(R_p\right) = \alpha^{-1} \Omega^{-1} \frac{R_p^2}{H^2} = 4\times
10^4\;\hbox{yrs}\;\alpha_{-1}^{-1} R_2^{3/2} \left(\frac{0.2
  R_p}{H}\right)^2\;,
\label{tII}
\end{equation}
where $\alpha_{-1} = \alpha/0.1$ is the disc viscosity parameter and $R_2
=R_p/100$AU. Type I migration is faster:
\begin{equation}
t_{\rm I}\left(R_p\right) = \Omega^{-1} \frac{M_*}{M_{\rm emb}}\frac{H}{R_p} =
3\times 10^3\;\hbox{yrs}\; R_2^{3/2} \frac{H}{0.2
  R_p} m_1^{-1}\;,
\label{tI}
\end{equation}
We can improve these estimates by building a simple
disc model that would specify how disc properties change with R.  We define
the stellar mass doubling time scale as $t_{\rm db} = M_*/\dot M$
where $\dot M$ is the mass accretion rate on the star, assumed to be constant
in time and radius inside the disc. By the order of magnitude, $t_{\rm db}$ should be
comparable to the free fall time of the host gaseous cloud from which the star
forms, e.g., $ \sim 10^5$ yrs for a 1 $\msun$ cloud at the typical interstellar
temperature of $10$ K \citep[e.g.,][]{Larson69}. 

Using the standard \cite{Shakura73} accretion disc formalism, we first find
$H(R)$ and $M_d(R)$ in the innermost disc. We then find the radius beyond
which the \cite{Toomre64} parameter falls below unity. In the outer
self-gravitating region of the disc we follow the treatment of
\cite{Goodman03} for a disc with no internal energy sources and enforce
$Q \approx (H/R)\;(M_*/M_d) = 1$. Finally, for a constant accretion rate disc,
$\dot M = 3 \pi \alpha H^2 \Omega \Sigma_d$, where $\Sigma_d \approx
M_d(R)/(\pi R^2)$ is the column depth of the disc, we can solve for
\begin{equation}
\frac{H}{R} = \left[\frac{1}{3\alpha \Omega t_{\rm db}}\right]^{1/3}\;.
\label{huponr}
\end{equation}
The accretion rate disc model built in this way is schematic but correctly
predicts the expected location of the self-gravitational region at $R_p >$ few
tens of AU to 100 AU. Using this disc model we are now able to calculate the
migration rate of a GE at an arbitrary $R_p$ for a given $t_{\rm db}$. We
follow the fit to numerical results by \cite{BateEtal03} to smoothly join the
type I and type II migration regimes. The fastest migration then occurs for embryo mass
$M_{\rm emb} = M_t$.


\subsection{Tidal disruption}\label{sec:disruption}

The Hill's radius of the embryo is (for $M_*=1 \msun$)
\begin{equation}
R_{\rm H} = R_p \left[\frac{M_{\rm emb}}{3M_*}\right]^{1/3}= \;0.15\; R_p
m_1^{1/3}\;. 
\label{r_hill}
\end{equation}
In general, radial migration accelerates as the GE moves in, whereas the
contraction of the envelope slows down with time (cf. equation
\ref{r_ass}). Therefore, as planets migrate inwards, there is a point
where the embryo's radius $R_{\rm emb}$ becomes comparable to the GE Hill's
radius. The tidal field from the parent star at this point starts to peel off
the outer layers of the embryo. The tidal radius, $R_t$, is defined as the
radius where $R_{\rm emb} = \eta_t R_{\rm H}$ ($\eta_t \simlt 1$). Using the
type II migration estimate as an example, we insert eq. \ref{tII} into
eq. \ref{r_ass}, and find
\begin{equation}
R_{\rm t} = 2.8 \;\hbox{AU}\; \eta_t^{-1} \alpha_{-1}^{1/2} R_2^{-3/4}
\kappa_{*}^{1/2} \frac{H}{0.2 R} \;m_1^{-1/3}\;.
\label{r_tidal}
\end{equation}
In Appendix we show that irradiation by the parent star can heat up the
outer layers of the envelope and even disrupt the whole GE. The effect appears
less important than tidal disruption for Solar type stars, but may become
dominant for higher mass stars.

\section{An illustrative example}\label{sec:example}

\begin{figure}
\centerline{\psfig{file=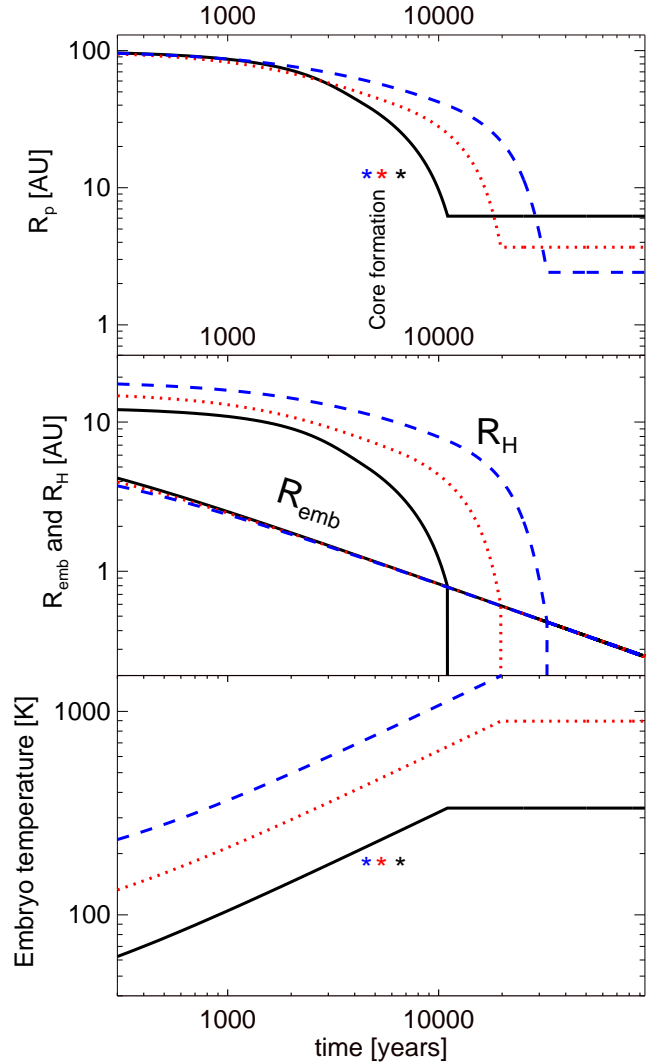,width=0.5\textwidth,angle=0}}
\caption{Evolution of GEs of mass 3, 6 and 10 Jupiter masses (black solid, red
  dotted and blue dashed, respectively) calculated as described in \S 2. {\bf
    Upper panel}: Radial position of the embryos versus time. The GEs start
  off at $R=100$ AU. The coloured asterisks and text ``core formation'' show
  $t_{\rm gr}$ -- the time when a solid core should form inside the embryo,
  with the colour matched to that of the curve. {\bf Middle}: Radius of the
  first core (power-law like curves), and the tidal radius at the current
  location of the first curve. The first core is assumed to be completely
  disrupted when the two sets of curves meet, i.e., when the tidal radius
  drops below $R_{\rm emb}$. {\bf Lower}: Temperature of the first cores as a
  function of time, with core formation times marked as above.}
\label{fig:3panel_migr}
\end{figure}

Figure \ref{fig:3panel_migr} shows the results of our giant embryo model
described in \S 2 for embryos of 3, 6 and 10 Jupiter masses, shown with
different colours. The parameters used for the figure are $t_{\rm db} = 10^5$
yrs, $\alpha=0.1$, $\kappa_0 = 0.01$, $M_* = 0.5 \msun$ (as the star is
assumed to be in the midst of its growth, rather than close to being
completely assembled), Solar metalicity, and the starting radial position is
$R_p = 100$ AU.

The upper panel shows radial migration of the embryos. Note that initially all
migrate at about the same rate (in the type II regime), but then the least
massive embryo overtakes the other two. The least massive one (black solid
curves) migrates the fastest as it happens to fall in the minimum of the
migration time curve, between type I and type II regimes \citep[see fig. 11
  in][]{BateEtal03}.  The asterisks and the text ``core formation'' mark the
time $t=t_{\rm gr}$, when the solid core is assumed to form in the centre of
the GEs (cf. eq. \ref{tgr}).

The middle panel shows the embryo size, $R_{\rm emb}$, and the respective
Hill's radius, for the three cases considered. We assume that when the two
sets of curves intersect, the embryo is disrupted, leaving only the core,
which migrates at a negligible rate during the calculation. This is why the
radial position curves, $R_p(t)$, approach a constant value in the upper panel
at late times. 

The lower panel shows the respective temperature evolution of the embryos
(cf. eq. \ref{T_of_t}). The least massive embryo forms the core last, as it is
the least dense of the three. However, it is also the coolest one. Due to
this, the least massive embryos may lock most of their ``metal'' content into
grains and actually yield heavier solid cores (see paper II). On the other
hand, the least massive embryo is disrupted sooner (in accord with
eq. \ref{r_tidal}), at $R_p \approx 6$ AU.

\section{The tidal downsizing hypothesis}

\subsection{Planets: leftovers of disrupted GE ?}\label{sec:left}

We found that for plausible conditions, the GE initial cooling time, the dust
growth time, the grain vaporisation time and the GE migration time are related
by
\begin{equation}
t_0 \ll t_{\rm gr} < t_{\rm migr} \sim t_{\rm vap}\;,
\label{tcond0}
\end{equation}
respectively. This order of timescales implies that (1) the GE contracts
quickly initially, which protects it from an immediate tidal disruption at the
point where it is born; (2) grains have enough time to grow and sediment to
the centre of the GE, forming a massive solid core there; (3) since the
cooling time becomes asymptotically long (\S \ref{sec:embryo}), the embryo is
tidally disrupted at the inner few AU.


\subsubsection{Terrestrial planets}\label{sec:terra}

If the migration time is shorter than the vaporisation time, 
\begin{equation}
t_{\rm migr} < t_{\rm vap}\;,
\label{tcond1}
\end{equation}
and the solid core mass, $M_{\rm core}$, is smaller than the critical core
mass, $M_{\rm cr}\sim 10 \mearth$ (see below), the gaseous component of the GE
is almost completely disrupted, e.g., disassociated from the solid core. As
the density of the solid cores is a few g cm$^{-3}$, the solid cores and any
smaller debris (if present) are not disrupted, and instead are released into
the protoplanetary disc around the GE location. The least massive of the
debris bodies may remain at that location in the disc for as long as $\simgt
10^6$ years, when the gaseous disc is presumably dissipated. After the
disruption, low mass solid cores are endowed with a tenuous gas atmosphere (paper II),
much of which may be lost thereafter.

\subsubsection{Giant planets with solid cores}\label{sec:jovian}

In paper II we argued that when the solid core grows more massive than a
critical core mass, $M_{\rm cr}$, the atmosphere near the core becomes so
dense that it is gravitationally unstable and starts to collapse on the solid
core. This ``core-assisted'' collapse is quite similar to that found in the
classical CA model \citep[e.g.,][]{Mizuno80,PollackEtal96}, except that gas is
accreted on the core from the GE rather than the disc. A proper numerical
calculation is required to find the exact value of the critical mass, $M_{\rm
  cr}$, but one can limit it from below by assuming an isothermal gaseous
envelope \citep[e.g., \S 6.1 in][]{Armitage10}, which yields $M_{\rm cr}
\simgt 3 \mearth$ (see paper II). The maximum value of $M_{\rm cr}$ can be
obtained by employing the classical CA model results -- the radiative zero
solution \citep{Stevenson82,IkomaEtal00}, and this yields $M_{\rm cr} \simlt
20 \mearth$.

We would thus posit that, if the tidal disruption of the GE occurs when the
core mass exceeds $M_{\rm cr}$, the outer metal-poor low density layers of the
GE are peeled off, but the inner high density solid core plus the bound gas
atmosphere remains.  The disruption leftover must be a metal rich (compared to
the parent star) giant planet, as are the giant planets in the Solar System.

\subsubsection{Giant planets without solid cores?}\label{sec:jovian}

The GE may go through the second wave of collapse when the temperature reaches
about 2000-2500 K \citep{Larson69,Bodenheimer74,Masunaga00} which can compact
the gas component to densities as high as $\sim$ g~cm$^{-3}$. The time to
reach that temperature is at least a few times the vaporisation time
(cf. eq. \ref{tvap}). Therefore, if
\begin{equation}
t_{migr} \gg t_{\rm vap}\;,
\label{tcond2}
\end{equation}
then the giant embryo may collapse as a whole. Detailed calculations of this
process in the planet formation context \citep[e.g.,][]{Bodenheimer74} show
that a Jupiter mass embryo only collapses in this way after $\sim 10^5$ years,
which appears longer than the migration time for ``reasonable''
parameters. However, more massive embryos start off hotter and should reach
the second collapse stage sooner (see paper II). Thus it is possible that
giant planets more massive than Jupiter may have no solid cores.


\section{Astrophysical implications}\label{sec:implications}


\paragraph*{When does planet formation stop?}
The view advanced in \S\S \ref{sec:terra} and \ref{sec:jovian} is an extreme
version of the tidal downsizing hypothesis, where everything ``interesting''
occurs inside the GEs before they are disrupted. This scenario is unlikely to
be realistic. When the GEs are dispersed, their remnants are still embedded in
a protoplanetary disc. The disc may be less massive, but as observations show
it may persist for a few million years. Therefore, it is likely that the
planet building action continues at that stage, now as in the CA model --
solid cores may grow further by accreting smaller solid debris; giant planets
may accrete more gas and also migrate radially. Therefore, the most reasonable
point of view is that the tidal downsizing hypothesis, if correct, may be a
new way to begin planetary system formation process \citep[see also][]{ClarkeLodato09}.

\paragraph*{Metalicity  correlation.} 
Observations show that the fraction of stars orbited by a giant planet
increases strongly with metalicity of the star \citep{FischerValenti05}. As
in the CA model, more solids in the gas enable more massive solid cores to be
built. We also assume that if no solid core is built, the GE is completely
destroyed close to the star; all of the GE material presumably accretes onto
the star.

In detail, core formation requires $t_{\rm gr} < t_{\rm vap}$ (\S
\ref{sec:dust}). Dust opacity, $\kappa_*$, is likely to be proportional to the
metalicity and the grain mass fraction, $f_g$. In such a model, the grain
growth timescale behaves as $t_{\rm gr}\propto f_g^{-5/21}$ (eq. \ref{tgr}),
whereas $t_{\rm vap} \propto f_g$. A higher $f_g$ thus gives the GE more time
to build a solid core. Further, \cite{Meru10} find that lower dust opacities
discs fragment closer to the star, reducing migration time
(cf. eq. \ref{tII}).  Low metalicity GEs are likely to vaporise their grains
before they make a solid core, and they also migrate inward too quickly to
their ultimate demise.

\paragraph*{Challenges and open questions.}\label{par:challenges}
``Tidal downsizing'' is a complex {\em hypothesis} that may turn out to not
work for at least the following reasons.  (1) giant planet embryos need to be
isolated or slowly accreting. If they pile up mass quickly by accretion from
the disc, they become too hot to support grain sedimentation (paper I).  (2)
We assumed that GEs migrate as planets do. This is probably a good
approximation when the radius of the embryo $R_{\rm emb}\ll R_{\rm H}$, but
much more work is needed to confirm this. (3) We have seen that the timescales
of the important processes (see \S \ref{sec:left}) do depend strongly on dust
opacity and other parameters. Real protoplanetary discs may be such that grain
growth inside the GEs is precluded or the embryos are disrupted too quickly.

\section{Conclusions}\label{sec:conclusion}

The tidal downsizing hypothesis combines some of the planet building processes
from both the CA and the GI models, preserving the strengths but not the
weaknesses of these theories. Open questions remain, but on the balance we
feel that the hypothesis deserves a further detailed consideration by the
planet formation community. We note that, having submitted this paper and
paper I, we discovered an independent suggestion on the possibility of
terrestrial core formation by tidally disrupted giant embryos by
\cite{BoleyEtal10}.

\section{Acknowledgments}

The author acknowledges illuminating discussions with and comments on the
draft by Richard Alexander, S.-H. Cha, Phil Armitage, Adam Burrows and Roman
Rafikov. Cathie Clarke, the referee, is thanked for suggestions that
significantly improved the clarity of the paper. Theoretical astrophysics
research at the University of Leicester is supported by a STFC Rolling grant.

\section*{Appendix: irradiative evaporation}

Irradiation of the GE by the star can affect the former strongly
\citep{CameronEtal82}. First we estimate the radius $R_{\rm lock}$ below which
the temperature of the surface layers of the GE is set by the irradiation
rather than the internal radiation. This is found by equating the stellar
radiation luminosity incident on the GE, $L_{\rm irr} = L_*\pi R_{\rm emb}^2/(4\pi
R_p^2)$, to the isolated GE radiative luminosity, $L_{\rm fc}(t)$. Since
$L_{\rm fc}(t) \propto T(t)^{-1} \propto (1 + 2t/t_0)^{-1/2}$ for the GE
(paper I), the result for $R_{\rm lock}$ at $t = t_{II}$ and $L_* = \lsun$ is
\begin{equation}
R_{\rm lock} \approx 12 \;\hbox{AU}\; m_1^{-1} k_*^{3/4}
\alpha_{-1}^{1/4} R_2^{-3/8}  \left(\frac{H}{0.2 R_p}\right)^{1/2}\;.
\label{rlock}
\end{equation}
Irradiation can also disrupt the GE when the irradiation temperature, $T_{\rm
  irr} = (L_{\rm irr}/4\pi R_{\rm emb}^2 \sigma)^{1/4}$ is comparable to the
mean temperature (equation \ref{T_of_t}) inside the GE, e.g., $T_{\rm irr} =
\eta_{ev} T(t)$, with $\eta_{ev} \simlt 1$. For one of the fixed planet
location runs \cite{CameronEtal82} found $\eta_{ev}\sim 0.5$ (cf. their
Fig. 2).  As $R_{\rm ev} = (L_*/4\pi \sigma T_0^4)^{1/2} (t_0/2t_{II})$, we
have
\begin{equation}
R_{\rm ev} \approx 1 \;\hbox{AU}\;  k_*
\alpha_{-1} R_2^{-3/2}  \left(\frac{H}{0.2 R_p}\right)^{2} \; \left({0.5 \over
  \eta_{ev}}\right)^2 \;m_1^{-1/3}
\label{r_ev}
\end{equation}
For fiducial parameters, tidal disruption of GEs occurs earlier (at larger
radii) than the irradiative evaporation. Furthermore, as noted by the referee,
the giant embryo may be shadowed by the disc if the disc scale-height, $H$, is
larger than the GE size, $R_{\rm emb}$. Thus we expect tidal disruption of the
GEs to be more frequent. Note however different parameter dependencies in
eq. \ref{r_tidal} and \ref{r_ev}. In general both processes may be important
in removing GE envelopes, with irradiation becoming more important for higher
mass stars.

\label{lastpage}

\end{document}